\documentclass[letterpaper, 10 pt, conference]{ieeeconf}  

\IEEEoverridecommandlockouts 

\usepackage{amsmath} 
\usepackage{tikz}
\usepackage{cite}

\title{\LARGE \bf
Estimating a Personalized Basal Insulin Dose from Short-Term Closed-Loop Data in Type 2 Diabetes
}

\author{Sarah Ellinor Engell$^{1,2}$, Tinna Björk Aradóttir$^{1}$, Tobias K. S. Ritschel$^{2}$,\\ Henrik Bengtsson$^{1}$, John Bagterp Jørgensen$^{2}$
\thanks{*This project is funded by Innovation Fund Denmark through the
Industrial PhD project 0153-00049B}
\thanks{$^{1}$Novo Nordisk A/S, DK-2880 Bagsværd, Denmark
        {\tt\small sqee@novonordisk.com, tbao@novonordisk.com, hbss@novonordisk.com}}%
\thanks{$^{2}$Department of Applied Mathematics and Computer Science,
        Technical University of Denmark, DK-2800 Kgs. Lyngby, Denmark
        {\tt\small saeg@dtu.dk, tobk@dtu.dk, jbjo@dtu.dk}}%
}

\begin{document}
\newcommand{\DrawIGInew}{
\begin{tikzpicture}[node distance = 2cm]
\tikzstyle{edge} = [->, thick, >=stealth];
\tikzstyle{dbarrow} = [thick, stealth-stealth];
\tikzstyle{vertex} = [circle, draw=black, thick, minimum size=1.2cm, fill=gray!20];
\tikzstyle{box} = [rectangle, draw=black, thick, minimum size=0.7cm];
\tikzstyle{vertex_me} = [circle, draw=black, thick, minimum size=1.2cm, fill=blue!10];
\tikzstyle{vertex_me2} = [circle, draw=black, thick, minimum size=1.2cm, fill=red!10];
\tikzstyle{subsystemB} = [rounded corners, blue, ultra thick];
\tikzstyle{subsystemBL} = [rounded corners, black, ultra thick];
\tikzstyle{subsystemY} = [rounded corners, yellow, ultra thick];

\node[vertex] (GA) at (-7, 1.2) {$G_{A}$};
\node[vertex, right of = GA] (GT) {$G_T$}; 
\node[vertex] (GP) at (-7, -1) {$G_p$};

\node[vertex_me2] (Gsc) at (-1.2, 3.2) {$G_{sc}$};

\node[vertex] (GC) at (-5, -1) {$G_{C}$};
\node[vertex, right of = GC] (GE2) {$G_{E2}$};
\node[box, right of = GE2] (Pan) {Pancreas};
\node[vertex] (Ieff) at (-3, -3) {$I_{E}$};
\node[vertex, right of = Ieff] (I) {$I$};
\node[vertex_me] (Iexo) at (-3, -5.75) {$I_{EXO}$};
\node[vertex_me] (S1F) at (-7, -5) {$S_{1,F}$};
\node[vertex_me, right of = S1F] (S2F) {$S_{2,F}$};
\node[vertex_me] (S1L) at (-7, -6.5) {$S_{1,L}$};
\node[vertex_me, right of = S1L] (S2L) {$S_{2,L}$};

\draw[edge] (GA) ++(-2,0) node[anchor = south west]{$d(t)$}-- (GA);
\draw[dbarrow] (GC) -- (GP);
\draw[edge, dashed] (GT) -| (Pan) node[above = 0.6cm, xshift =-0.3cm]{$+$} ;
\draw[edge] (GA) -- (GT);
\draw[edge] (GT) -- (GC);
\draw[edge] (GC) ++(-1,1) node[anchor=south,xshift=0.1cm]{$EGP$}-- (GC);
\draw[edge] (GE2) -- +(0.8,-0.8) node[anchor = south, yshift = -0.4cm] {$k_{GE2}$};
\draw[edge] (GC) -- +(0.8, -0.8)node[below=0.1cm] {$CL_{G}$};
\draw[edge] (GC) -- +(0, -2)node[below=0.1cm] {$CL_{GI}$};
\draw[edge] (Gsc) --  +(2.2, 0) node[anchor=south east]{$y_{cgm}(t)$};
\draw[edge,dashed,color=red] (GC) -- (Gsc);
\draw[edge, dashed] (GC) -- (GE2);
\draw[edge, dashed] (GE2) -- (Pan) node[above=0.1cm, xshift=-1.2cm]{$+$};
\draw[edge] (Pan) -- (I);
\draw[edge, dashed] (I) -- (Ieff); 

\draw[edge] (S1F) ++(-2, 0) node[anchor = south west]{$u_{F}(t)$} -- (S1F);
\draw[edge] (S1F) --(S2F);
\draw[edge] (S1L) ++(-2, 0) node[anchor = south west]{$u_{L}(t)$} -- (S1L);
\draw[edge] (S1L) -- (S2L);
\draw[edge] (S2F) -- (Iexo);
\draw[edge] (S2L) -- (Iexo);
\draw[edge] (Iexo) -- +(0.8, -0.8) node[anchor=south west] {$k_{exo}$};
\draw[edge, dashed, color=blue] (Iexo) -- (Ieff);
\draw[edge, dashed] (Ieff) -- +(-1.5, 0) node[above=0.1] {$+$};
\draw[edge] (Ieff) -- +(0.8, -0.8) node[anchor=south, xshift =0.2cm] {$k_{IE}$};
\draw[edge] (I) -- +(0.8, -0.8) node[anchor=south, xshift =0.2cm]{$CL_{I}$};

\path (GA) ++(-1, 1) coordinate (c1);
\path (GT) ++(0.5, 1) coordinate (c2);
\path (Pan) ++(1.5, 0.5) coordinate (c3);
\path (I) ++(1.5, -1) coordinate (c4);
\path (I) ++(-7, -1) coordinate (c5);

\draw[subsystemBL] (c1) rectangle (c4);

\end{tikzpicture}
}

\newcommand\copyrightnotice{
    \begin{tikzpicture}[remember picture,overlay]
    \node[anchor=south,yshift=10pt] at (current page.south) {\parbox{\dimexpr\textwidth}{\footnotesize\copyright2022 IEEE.  Personal use of this material is permitted. Permission from IEEE must be obtained for all other uses, in any current or future media, including reprinting/republishing this material for advertising or promotional purposes, creating new collective works, for resale or redistribution to servers or lists, or reuse of any copyrighted component of this work in other works.}};
    \end{tikzpicture}
}

\maketitle
\thispagestyle{empty}
\pagestyle{empty}

\begin{abstract}
In type 2 diabetes (T2D) treatment, finding a safe and effective basal insulin dose is a challenge. The dose-response is highly individual and to ensure safety, people with T2D “titrate” by slowly increasing the daily insulin dose to meet treatment targets. This titration can take months. To ease and accelerate the process, we use short-term artificial pancreas (AP) treatment tailored for initial titration and apply it as a diagnostic tool. Specifically, we present a method to automatically estimate a personalized daily dose of basal insulin from closed-loop data collected with an AP. Based on AP-data from a stochastic simulation model, we employ the continuous-discrete extended Kalman filter and a maximum likelihood approach to estimate parameters in a simple dose-response model for 100 virtual people. With the identified model, we compute a daily dose of basal insulin to meet treatment targets for each individual. We test the personalized dose and evaluate the treatment outcomes against clinical reference values. In the tested simulation setup, the proposed method is feasible. However, more extensive tests will reveal whether it can be deemed safe for clinical implementation.
\newline
\end{abstract}

\section{Introduction}
\copyrightnotice
Worldwide, one in eleven people live with diabetes, whereof approximately 90\% have type 2 diabetes (T2D). Left untreated, people with T2D suffer from persistent high blood glucose levels that eventually lead to complications in many parts of the body. 
Fortunately, numerous treatment options exist. As T2D progresses, daily injections of basal insulin  become necessary to lower the elevated blood glucose levels \cite{Draznin2022}. However, basal insulin initiation, a process known as titration, is challenging as the insulin response in the body varies greatly between individuals. It is crucial to avoid overdosing as too much insulin can quickly cause life-threatening low glucose levels. To obtain a safe and effective dose, the amount of injected insulin is gradually increased in size, until the desired fasting blood glucose level is reached. The insulin dose is adjusted manually based on pre-breakfast finger-prick blood glucose measurements. Typically, this titration is performed at home and can take several months. For more than half of the individuals initiating insulin treatment, the task is so demanding that it leads to non-adherence and failed insulin titration \cite{Khunti2020}. In the future, the burden of self-titration may be overcome with automated titration solutions.

Several automated solutions have been proposed in the literature, ranging from model-free extremum seeking control \cite{Krishnamoorthy2021}, to model predictive control \cite{Aradottir2019}, and iterative learning \cite{Cescon2021}. In simulation, these methods have shown to speed up the titration process, improve safety and reduce the workload compared to standard-of-care methods. A few methods have been tested in clinical trials with promising results \cite{Aradottir2020,Donnelly2015,MIDS2020}. Still, simple self-titration remains the standard-of-care solution in clinics today \cite{Draznin2022}.

Another way to automate insulin treatment is through closed-loop control with an artificial pancreas (AP) system. In recent years, this has become a viable treatment option for people with type 1 diabetes \cite{Boughton2019}. In the coming years, commercial AP systems are expected to become available to people with T2D as well \cite{peters2022a}. An AP system consists of a control algorithm that, based on frequent sensor measurements from a continuous glucose monitor (CGM), automatically adjusts and infuses fast-acting insulin via an insulin pump to achieve target glucose values. Although these systems automate insulin dose selection and delivery, their technical complexity may limit the uptake in an older T2D population \cite{Reiterer2018}. In light of this, the greater population’s treatment needs may be met with simpler injection-based solutions. However, the emergence of closed-loop treatment for T2D can enable new forms of automated titration through short-term AP-use \cite{Engell2021}. We propose that pump-induced system excitation 
can determine what basal insulin dose will bring each individual to treatment targets on once-daily injection-based treatment.

In this work, we present a method to estimate a personalized basal insulin dose from short-term closed-loop data. Based on data from a stochastic simulation model, we use maximum likelihood estimation (MLE) to identify parameters in a simpler prediction model for 100 virtual people. For a given set of parameters, we use the continuous-discrete extended Kalman filter (CDEKF) to approximate the likelihood function which is maximized in MLE. With the identified model, we compute a personalized insulin dose to meet treatment targets. Finally, we test the computed daily dose of insulin in our simulation model and evaluate the treatment outcomes.

This paper is organized as follows. In Section \ref{sec:methods}, we present the two physiological models for data generation and parameter estimation. We briefly describe the parameter estimation technique. Section \ref{sec:results} presents the results with the proposed method for three different data-collection scenarios. In Section \ref{sec:discussion}, we evaluate and discuss the performance of the tested control strategy. Section \ref{sec:conclusion} concludes the paper and presents ideas for future work.


\section{Methods}
\label{sec:methods}

In this section, we introduce the simulation model used to generate data for 100 virtual people on closed-loop treatment. We use the data for parameter estimation in a simpler prediction model, presented in Section \ref{sec:predict}. We use the CDEKF and MLE to identify model parameters. Section \ref{sec:CDEKF} and \ref{sec:MLE} briefly describe the estimation technique. To conclude, we present how an optimal basal insulin dose is calculated from the estimated parameters.


\subsection{Simulation Model}
\begin{figure}[tb]
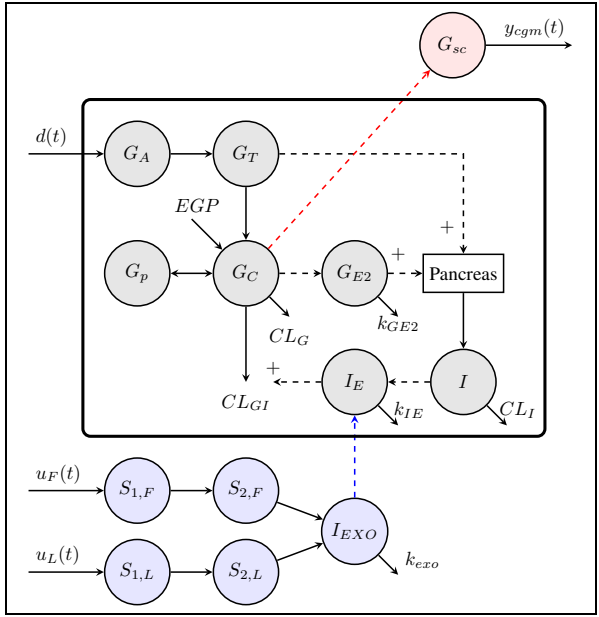

    \centering
    \framebox{\parbox{3in}{\resizebox{0.87\columnwidth}{!}{
    \DrawIGInew
    }}}
    \caption{Model Structure for the Simulation Model. Meals, $d(t)$, fast-acting insulin, $u_F(t)$, and long-acting insulin, $u_L(t)$, are the inputs. The continuous glucose monitor (CGM) outputs the subcutaneous glucose concentration, $y_{cgm}(t)$. The compartments denote the glucose absorption, $G_A$, glucose transport, $G_T$, peripheral glucose, $G_p$, central glucose, $G_C$, subcutaneous glucose, $G_{sc}$, glucose effect on insulin secretion, $G_{E2}$, plasma insulin, $I$, insulin effect, $I_E$, and the exogenous insulin, $I_{EXO}$. Exogenous fast-acting insulin is absorbed via the compartments $S_{1,F}$ and $S_{2,F}$, and exogenous long-acting insulin is absorbed through $S_{1,L}$ and $S_{2,L}$. The shown inputs and outputs from compartments are endogenous glucose production, $EGP$, glucose-dependent clearance, $CL_G$, insulin-dependent clearance, $CL_{GI}$, endogenous insulin clearance, $CL_I$, exogenous insulin clearance rate, $k_{exo}$, and the rate constants for effect delay, $k_{IE}$ and $k_{GE2}$.}
    \label{fig:IGI}
\end{figure}

To simulate a cohort of 100 virtual people with T2D, we employ a stochastic version of the integrated glucose-insulin (IGI) model \cite{Roge2014,Engell2021}. The model consists of 14 differential equations that together describe how glucose and insulin interact in the human body. We apply an extended version where exogenous fast- and long-acting insulin can be added as inputs and the subcutaneous blood glucose can be measured. Fig. \ref{fig:IGI} shows the model structure and the model equations are listed in \cite{Engell2021}.

The glucose-insulin dynamics are a continuous process observed through discrete measurements,
\begin{subequations}
    \begin{align}
        dx(t) &= f(t,x(t),u(t),d(t),\theta) dt + \sigma d\omega(t) \\
        y_k &= h(t_k,x(t_k)) + v_k
    \end{align}
\end{subequations}
where $x(t)$ is the state vector, $u(t)$ is the input vector containing both $u_F$ and $u_L$, $d(t)$ is the meal disturbance, and $\theta$ constitutes the model parameters. The drift function, $f$, is given by the IGI model. For the input, we assume a zero-order hold parametrization, i.e. $u(t) = u_k$ for $t_k \leq t < t_{k+1}$. The process noise, \{$\omega(t), t\geq 0$\}, is a standard Wiener process and its increment has covariance $Idt$. $\omega(t)$ is scaled by a time-invariant diagonal matrix, $\sigma$, adding noise to the central glucose compartment, $G_c$. The measurement noise on $y_{cgm}$ is assumed normally distributed, $v_k \sim N_{iid}(0,R_k)$.

\subsection{Prediction Model}
\label{sec:predict}
We use the fasting blood glucose model by Aradóttir et al. \cite{Aradottir2018} to obtain a personalized dose-response model for each virtual person. The authors designed the model such that it allows for identification of glucose-insulin dynamics with one input (insulin) and one output (fasting glucose) \cite{Aradottir2018}. The model consists of four differential equations,
\begin{subequations}
\begin{align}
    \frac{dx_1(t)}{dt} &= \frac{1}{p_1}u(t) - \frac{1}{p_1}x_1(t)\\
    \frac{dx_2(t)}{dt} &= \frac{1}{p_1}x_1(t) - \frac{1}{p_1}x_2(t)\\
    \frac{dx_3(t)}{dt} &= p_3(x_2(t) + p_7x_4(t)) - p_3x_3(t)\\
    \frac{dx_4(t)}{dt} &= -(p_5 + p_4x_3(t))\cdot x_4(t) + p_6,
\end{align}
\label{eq:predict}
\end{subequations}
that represent the glucose-insulin dynamics in a human body. The states $x_1$ [U/min] and $x_2$ [U/min] describe the body's absorption of insulin input, $u$ [U/min]. The effect of the insulin is represented by $x_3$ [U/min] and the blood glucose concentration is $x_4$ [mmol/L]. The system outputs discrete sensor measurements,
\begin{equation}
    y_k = x_{4}(t_k) + v_k.
\end{equation} 
As in the simulation model, the measurement noise is assumed normally distributed, $v_k \sim N_{iid}(0,R_k)$.

We estimate the parameters, $\theta =$ [$p_4$; $p_6$; $p_7$], as these are known to be identifiable from sparse data \cite{Aradottir2018} and therefore may also be identified from our intense data capture. We apply published population parameters for $p_1$, $p_3$ and $p_5$. We use the published population parameters as the initial guess for $p_4$, $p_6$, and $p_7$ in the parameter estimation. Parameter descriptions and published values are found in Table \ref{tb_par}.

\begin{table*}[ht]
\caption{Population Parameters for the Prediction Model}
\label{tb_par}
\begin{center}
\begin{tabular}{c l l l c}
\hline
Parameter & Value & Unit & Description & Reference\\
\hline
$p_1$ & $60$ & [min]  & Time constant for fast-acting insulin absorption & \cite{Kanderian2009} \\
$p_3$ & $0.011$ & [1/min]  & Delay in insulin action & \cite{Aradottir2018} \\
$p_4$ & $0.44$ & [1/U]  & Insulin sensitivity & \cite{Aradottir2018}\\
$p_5$ & $0.0023$ & [1/min]  & Insulin-independent glucose clearance & \cite{Aradottir2018}\\
$p_6$ & $0.0672$ & [mmol/L$\cdot$min]  & Endogenous glucose production & \cite{Aradottir2018}\\
$p_7$ & $0.0018$ & [U$\cdot$L/mmol$\cdot$min] & Endogenous insulin production & \cite{Aradottir2018}\\
\hline
\end{tabular}
\end{center}
\end{table*}

\subsection{Data Generation}
To simulate a cohort of a hundred virtual patients, we draw parameters from the published distribution for the insulin sensitivity and insulin production \cite{Roge2014}. We select body weights from the distribution in \cite{Zinman2012} and scale the weight-dependent parameters accordingly. After parameter selection, we screen the virtual people to ensure that their insulin response is feasible for a T2D population. Before insulin treatment, $95\%$ of the individuals in \cite{Zinman2012} have a fasting blood glucose level below 15 mmol/L. As the cohort in \cite{Zinman2012} is a subset of the insulin-requiring T2D population in the real world, we allow for higher fasting blood glucose values in our simulated cohort. When no insulin is given, the fasting blood glucose must lie within a 7.5-20 mmol/L range. Additionally, the insulin dose required to reach a glucose level of $5.8$ mmol/L must not surpass $150$ U. If the constraints are violated, we re-sample the model parameters until the constraints are met.

As a simplified AP system, we employ an integrator-based control algorithm \cite{Engell2021} that drives the blood glucose towards the 5.8 mmol/L reference value. We simulate closed-loop treatment in a fasting state with no meals, $d(t) = 0$, for 24 and 48 hours. The selected scenario does not represent a realistic setup to apply in clinic. However, it facilitates an undisturbed assessment of how the controller gain and the duration of excitation influences the quality of a target dose estimate for basal insulin.

To mimic the continuous-discrete nature of sensor measurements from a physiological system, we simulate the IGI model using an Euler-Maruyama scheme with a time step size of one minute. Every five minutes, the CGM outputs a noise-corrupted measurement, $y_k$, of the subcutaneous glucose concentration. When estimating parameters in the prediction model, we use the CGM measurements from this simulation as input to the CDEKF.

The simulation and parameter estimation was implemented in \texttt{Matlab R2020b}.

\subsection{Continuous-Discrete Extended Kalman Filter}
\label{sec:CDEKF}
We use the iterative framework of the CDEKF for parameter estimation. At every sample point, $k$, we update the estimate of our system states, $\hat{x}_{k|k-1}$, and the state covariance matrix, $P_{k|k-1}$, using the incoming measurement, $y_k$. For this update, we compute the innovation,
\begin{equation}
    e_k = y_k - \hat{y}_{k|k-1}
\end{equation}
as the difference between the measured value, $y_k$, and the model predicted output, $\hat{y}_{k|k-1} = C_k\hat{x}_{k|k-1}$. The matrix $C_k$ is a linearization of the measurement equation, $h(t_k,\hat{x}_{k|k-1})$, at the current state estimate, $\hat{x}_{k|k-1}$,
\begin{equation}
    C_k = \frac{\partial h}{\partial x} (t_k,\hat{x}_{k|k-1}).
\end{equation}
Using the variance of the measurement noise, $R_k$, we can obtain the covariance of the innovation signal, $R_{e,k}$, and compute the Kalman gain, $K_k$,
\begin{subequations}
    \begin{align}
        R_{e,k} &= C_kP_{k|k-1}C_k^T + R_k,\\
        K_k &= P_{k|k-1}C_k^TR_{e,k}^{-1}.
    \end{align}
\end{subequations}
Finally, we update the estimate of the states and their covariance,
\begin{subequations}
    \begin{align}
    \begin{split}
        \hat{x}_{k|k} &= \hat{x}_{k|k-1} + K_ke_k,\\
        P_{k|k} &= (I-K_kC_k)P_{k|k-1}(I-K_kC_k)^T \\
        & \qquad + K_kR_kK_k^T.
    \end{split}
    \end{align}
\end{subequations}
To obtain the one-step prediction of the states and their covariance, we solve a system of differential equations,
\begin{subequations}
    \begin{align}
        \frac{d\hat{x}_k(t)}{dt} &= f(t,\hat{x}_k(t),u_k, d_k, \theta),\\
        \frac{dP_k(t)}{dt} &= A_k(t)P_k(t) + P_k(t)A_k(t)^T + \sigma \sigma^T,
    \end{align}
\end{subequations}
with the initial conditions
\begin{subequations}
    \begin{align}
        \hat{x}_k(t_k) &= \hat{x}_{k|k},\\
        P_k(t_k) &= P_{k|k},
    \end{align}
\end{subequations}
and where
\begin{equation}
\begin{split}
    A_k(t) &= A(t,\hat{x}_k(t),u_k, d_k, \theta)\\ 
    &= \frac{\partial f}{\partial x}(t,\hat{x}_k(t),u_k,d_k,\theta)
    \end{split}
\end{equation}
is a linearization of the drift function $f$ evaluated at $\hat{x}_{k}(t)$ with input $u_k$, disturbance $d_k$, and parameters $\theta$.

\subsection{Maximum Likelihood Estimation}
\label{sec:MLE}
From a discrete series of measurements,

\begin{equation}
    \mathcal{Y}_N = \{y_0,y_1,...,y_N\},
\end{equation}
obtained from the simulation model, we estimate the parameter set, $\theta$, that maximizes the conditional probability,

\begin{equation}
    p(\mathcal{Y}_N|\theta) = p(y_N,y_{N-1},...,y_0|\theta).
    \label{eq_jpd}
\end{equation}

This is equivalent to minimizing the negative log-likelihood as a function of $\theta$, i.e.
\begin{equation}
    \hat{\theta} = \underset{\theta}{\arg\min} ~ V(\theta)
\end{equation}
where
\begin{align}
\begin{split}
    V(\theta) &= -\ln(p(\mathcal{Y}_N|\theta))\\
    &= \frac{1}{2}(N+1)n_y \ln(2\pi)\\
    & \qquad + \frac{1}{2} \sum_{k=0}^N \ln[\det(R_{e,k})] + e_k^TR_{e,k}^{-1}e_k.
        \label{eq_Vtheta}
\end{split}
\end{align}
Here, $e_k$ and $R_{e,k}$ are CDEKF outputs for a selected parameter set $\theta$. $n_y$ denotes the number of system outputs.

\subsection{Computing the Target Insulin Dose}
Once we identify a set of parameters for a personalized dose-response model, we calculate a daily insulin dose. With the estimated parameter set, we solve for the insulin infusion rate in (\ref{eq:predict}),

\begin{equation}
    u_{target} = \frac{p_6 - y_{ref}\cdot p_5}{y_{ref} \cdot p_4} - p_7\cdot y_{ref}
\end{equation}
that will bring the blood glucose concentration to the desired reference value, $y_{ref}=5.8$ mmol/L. The infusion rate, $u_{target}$, is given in U/min. To get a daily dose, we calculate the total insulin delivered over 24 hours,
\begin{equation}
    u_{basal} = u_{target} \text{ [U/min]} \cdot 60 \text{ [min/h]} \cdot 24 \text{ [h/day]}
\end{equation}
In our simulation model, we inject the daily dose of basal insulin, $u_{basal}$, at 7:00 AM on the five consecutive days after closed-loop treatment.






\section{Results}
\label{sec:results}

\begin{figure}[tb]
\centering
\includegraphics[width=\columnwidth,trim=0 1.5cm 1cm 1cm, clip]{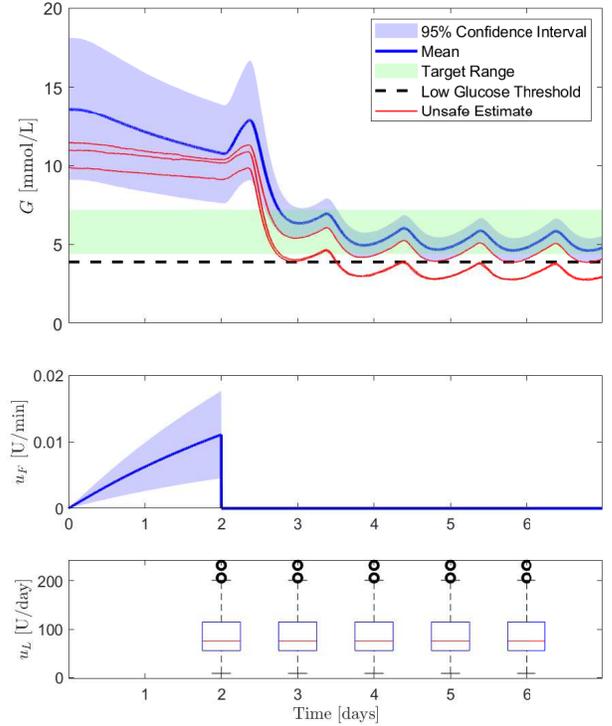}
\caption{48 Hours of Closed-Loop Data for 100 Virtual People. In the closed-loop period, glucose levels, $G$, are driven towards the $4.4-7.2$ mmol/L target range by fast-acting insulin infusion, $u_F$. Based on the recorded closed-loop data, a target insulin dose is computed and administered as a daily injection of long-acting insulin, $u_L$, in the five last simulation days. In red, we plot the individual curves where the glucose level drops below $3.9$ mmol/L.}
\label{fig:48}
\end{figure}

In the first simulation scenario, we collect closed-loop data for 48 hours as shown in Fig. \ref{fig:48}. 
Throughout the closed-loop period, the controller gradually increases the infused insulin and the glucose levels are steered towards the green target area for the 100 virtual people. Based on the collected data, we compute a basal insulin dose at the end of day 2 and implement it on day 3. After the switch to injection-based treatment on day 3, the majority of the simulated people have glucose levels within the 4.4-7.2 mmol/L target area. For three virtual people, the calculated insulin dose is too high and the glucose levels drop below 3.9 mmol/L. This is dangerously low, and would not be accepted in a clinical implementation. Note that the poor dose estimates do not coincide with the outliers in the boxplot of basal insulin doses. The three virtual people with poor insulin dose estimates show a minimal reduction in glucose values during the closed-loop period. We expect that a higher system excitation for these individuals, e.g. a more aggressive controller, can improve dose estimates.

\begin{figure}[tb]
\centering
\includegraphics[width=\columnwidth,trim=0 1.5cm 1cm 1cm, clip]{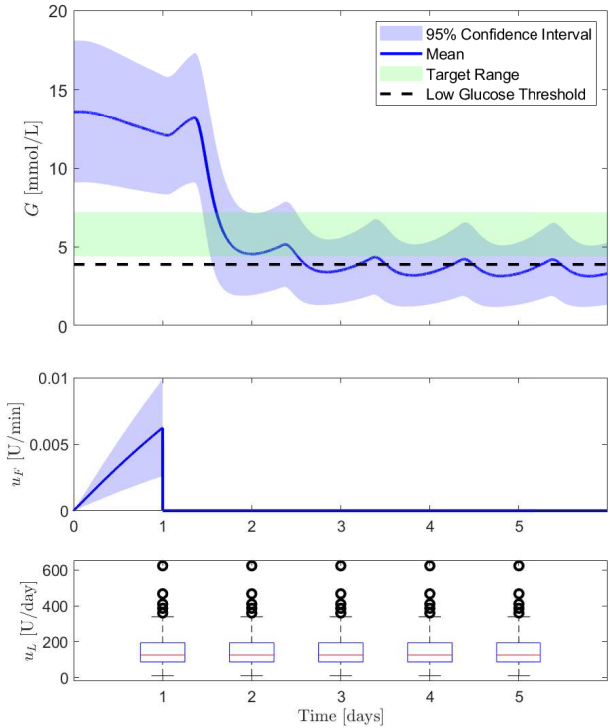}
\caption{24 Hours of Closed-Loop Data for 100 Virtual People. For 24 hours, the control algorithm gradually increases fast-acting insulin, $u_F$, to steer the blood glucose, $G$, into the $4.4-7.2$ mmol/L target range. After the closed-loop data collection, we estimate a personalized, daily insulin dose. We simulate the outcomes when the dose is administered as a daily injection of long-acting insulin, $u_L$. With the short data-collection period, we overestimate the required daily dose of insulin for 78 people.}
\label{fig:24}
\end{figure}

Across the simulated cohort, the general performance is good when 48 hours of data is used to estimate a personalized basal insulin dose in a fasting scenario. We wish to determine whether an equivalent performance can be reached with less data. In Fig. \ref{fig:24}, we see the outcomes for only 24 hours of closed-loop data collection.

\begin{figure}[tb]
\centering
\includegraphics[width=\columnwidth,trim=0 1.5cm 1cm 1cm, clip]{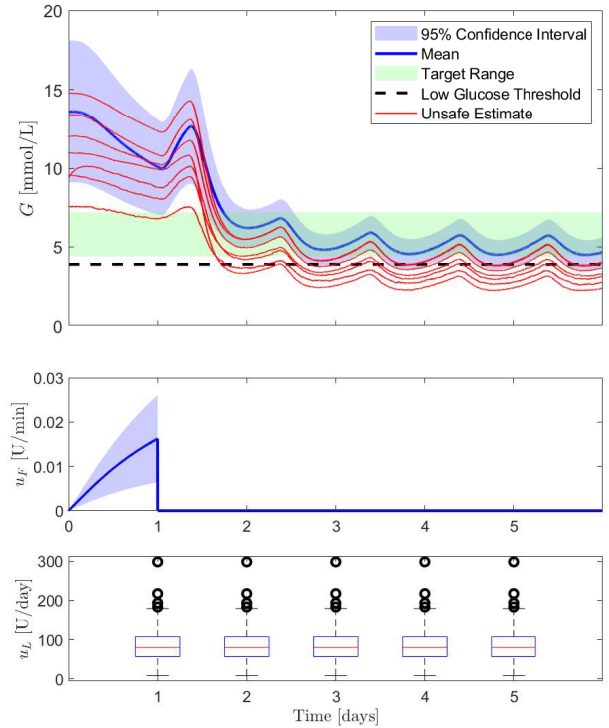}
\caption{Tripled Controller Gain and 24 Hours of Closed-Loop Data. To improve the system excitation, we triple the controller gain compared to Fig.~\ref{fig:24}. As a result, the blood glucose, $G$, drops quicker towards the $4.4-7.2$ mmol/L target range, and the pump infuses more fast-acting insulin, $u_F$. We see that the daily dose estimate of long-acting insulin, $u_L$, is safer. Only seven people experience blood glucose concentrations below $3.9$ mmol/L. In red, we show the seven individual curves with poor dose estimates.}
\label{fig:24_3Ki}
\end{figure}

With 24 hours of data, 78\% of the basal insulin doses are overestimated, driving blood glucose concentrations far below the 3.9 mmol/L threshold. In conclusion, the system excitation does not appear to be sufficient to capture essential system dynamics. In an attempt to increase system excitation and improve performance, we increase the controller gain by a factor of three. The result is shown in Fig. \ref{fig:24_3Ki}.

With a tripled controller gain over a 24-hour period, we see an improved performance compared to the nominal gain. Of the 100 virtual people, only seven have overestimated doses. As in the 48 hour simulation, the people with poor dose estimates have a smaller gradient compared to the population mean. This could indicate a lower degree of system excitation with the chosen controller gain. The majority of simulated people achieve target glucose values when treated with the computed basal insulin dose. Still, the best performance is seen in the scenario where closed-loop data is collected over 48 hours, suggesting that both data quantity and system excitation are crucial if this method is to be applicable in clinical practice.

\section{Discussion}
\label{sec:discussion}
In this work, we investigate the feasibility of an automated titration solution for people with T2D. We show how a closed-loop system may be used for system excitation and enable target dose estimation. Both the the magnitude of the controller gain and the length of the closed-loop treatment affect the efficacy and safety of the proposed method. 

The controller gain applied in Fig. \ref{fig:48} and \ref{fig:24}, results in a total daily dose of less than $0.2$ U/kg body weight after 24 hours. This is in accordance with standard-of-care titration guidelines for basal insulin that recommend an initial daily dose of 0.1-0.2 U/kg body weight. We have seen in Fig. \ref{fig:24_3Ki} that an increased controller gain excites the underlying system to a greater extent, and consequently, the parameter estimates improve. In a real-world setting, an increased controller gain may not cause a direct risk of low blood glucose levels, however, a sudden drop in glucose concentration driven by the AP can be highly uncomfortable for the user. Additionally, people with sustained high blood glucose levels over long periods are at risk of nerve and eye damage when blood glucose decreases rapidly \cite{Gibbons2020}. 

Ideally, the AP system should be worn for several days with a moderate gain to estimate a safe and effective basal insulin dose. Modern patch pumps, i.e. tubeless insulin pumps that are fixed to the skin with adhesives, have a wear-time of 72 hours. These pumps could provide a user with a convenient way to collect multiple days of data for system identification. However, if the user is expected to refrain from eating in the whole closed-loop period, the parameter estimation must be made feasible within a shorter time frame, e.g. 12 hours. This may not be possible. In simulation, we can choose to disregard multiple disturbances from meals and interday variations in insulin response. In reality, the identification is more complicated. In an uncontrolled real-world setting, the complexity of the model identification process will increase as glucose excursions after, e.g. undocumented meals interfere with the administered insulin infusion rate. A way to circumvent this could be to introduce controlled meal tests, i.e. known quantities of carbohydrates consumed at fixed hours. The meal tests can be used for additional system excitation and will additionally make the identification process more comfortable for the user. To improve system identification, the controller input and meal tests could be tuned in an optimal design of experiment.

In a real-world setting, we may experience the unfortunate situation that the model parameters cannot be estimated from the collected data. If no more closed-loop data collection is possible, we propose a unit-to-unit conversion from the pump infusion rate to an injection-based dose of basal insulin, followed by manual titration. In this way, the titration already performed by the AP would not be lost.

In the case where dose estimates are found, clinicians can be hesitant to deem them safe. To increase the safety margin in a clinical implementation, a fraction of the predicted dose may be used instead of the full dose, e.g. 75\% of the predicted dose. Alternatively, the daily injection size can be increased in controlled steps until the predicted target dose is reached. Compared to standard-of-care titration, these steps would be larger and would allow us to reach treatment targets faster. A step-wise increase in dose size may be safer and less unpleasant for the user, as it will result in a more controlled decrease in blood glucose.
Another way to test the predicted dose would be to continue the pump treatment and increase the infusion rate. In this way, it remains possible to quickly shut off insulin infusion if the predicted dose brings the blood glucose into dangerously low values.

In this paper, we test the proposed method on a simple simulated scenario. As a result, the implementation is, in the current state, not ready for clinical use. However, this work presents a new approach to insulin titration in T2D that may hold clinical potential. For future work, a higher complexity in the simulation scenario will allow evaluation in a setup that closer resembles real-world cases.

\section{Conclusion}
\label{sec:conclusion}
In this work, we employ closed-loop data for system identification in people with T2D. Based on 24-48 hours of glucose-insulin data, we identify a personalized basal insulin dose using the CDEKF and maximum likelihood estimation. The proposed method is feasible in the chosen simulation setup, however, the efficacy and safety of the dose estimates heavily depend on the system excitation. We can affect the system excitation by increasing the controller gain and extending the data collection period. In future work, we aim to instigate how meal tests can be included in using optimal experiment design to make the implementation viable in a real-world setting allowing people to eat.


\addtolength{\textheight}{-12cm}   








\bibliography{CDC.bib}
\bibliographystyle{IEEEtran}

\end{document}